# Experimental setup for measuring the barocaloric effect in polymers: Application to natural rubber

N. M. Bom[1], E. O. Usuda[1], G. M. Guimarães[1], A. A. Coelho[2] and A. M. G. Carvalho[1]

[1]*Laboratório Nacional de Luz Síncrotron, CNPEM, 13083-100 Campinas, Brazil*
[2]*Instituto de Física Gleb Wataghin, UNICAMP, 13083-970 Campinas, Brazil*

Barocaloric materials have shown to be promising alternatives to the conventional vapor-compression refrigeration technologies. Nevertheless, barocaloric effect ($\sigma_b$-CE) has not been extensively examined for many classes of materials up to now. Aiming at fulfilling this gap, this paper describes the development of a high-pressure experimental setup for measuring the $\sigma_b$-CE in polymers. The design allows simultaneous measurements of temperature, pressure and strain during the barocaloric cycle. The system proved to be fully-functional through basic experiments using natural rubber. Samples exhibited large temperature variations associated to the $\sigma_b$-CE. Strain-temperature curves were also obtained, which could allow indirect measurements of the isothermal entropy change.

## INTRODUCTION

The current requirements for energy-saving and eco-friendliness lead to the development of substitutes for the conventional cooling devices. Solid-state cooling based on *i*-caloric effects appeared as a promising solution. The *i*-caloric effects correspond to an isothermal entropy change or an adiabatic temperature change in a material upon the application of an external field (magnetic, electric, mechanical stress). The "*i*" stands for intensive thermodynamic variables. They can be categorized as magnetocaloric effect (*h*-CE), electrocaloric effect (*e*-CE) and mechanocaloric effect ($\sigma$-CE). $\sigma$-CE is subdivided in: i) elastocaloric ($\sigma_e$-CE), driven by uniaxial stress; and ii) barocaloric ($\sigma_b$-CE), driven by isostatic pressure. Some important issues still hinder the usage of *h*-CE and *e*-CE in commercial devices. Magnetocaloric materials are based on rare-earth compounds,[1] which are very costly and harmful for the environment.[2] Moreover, large magnetic fields (2-5 T) are required to induce significant temperature variations.[3–5] For electrocaloric materials, there is a risk of breakdown when high electric fields are applied.[6,7] Besides, only thin film electrocaloric materials exhibit significant changes in temperature.[6]

Mechanocaloric materials have attracted the attention in the last years[8]. Giant $\sigma_e$-CE and $\sigma_b$-CE have been demonstrated in shape memory alloys (SMAs).[9,10] Nevertheless, the relatively large stresses/pressures required to induce $\sigma$-CE in SMAs may represent a drawback for cooling applications. On the other hand, two families of polymers show promising mechanocaloric potential: PVDF-based polymers and elastomers. Among elastomers, natural rubber (NR) is prominent material due to its sustainability, low-cost and high caloric performance. Barocaloric materials represent fertile ground for mechanocaloric research, since they exhibit high caloric efficiency.[5] Despite this, there is still a lack of literature on $\sigma_b$-CE for many classes of materials,[11] such as the polymers. Therefore, the goal of the present study is to develop and optimize an experimental setup for investigation of $\sigma_b$-CE in polymers. The system is composed of a high-pressure chamber and a data acquisition set, being capable of controlling and measuring the relevant process variables: temperature, pressure and strain. Tests demonstrate the functionality of the setup. NR samples exhibit a similar behavior in comparison with analogous mechanocaloric studies.

## EXPERIMENTAL SETUP

The constructed apparatus (Figure 1) consists of three main parts: i) a cylindrical chamber with a through-hole (12 mm diameter) in the center; ii) a piston; and iii) a bottom closure. Underneath this set, a cylindrical slab supports the chamber and the closure. All referred pieces were made of stainless steel, in order to reduce the heat transfer. We also built all the pieces with carbon steel, getting the same results. A polymer sample is placed at the center of the device, between the piston and the closure. Surrounding the chamber, a copper coil enables the circulation of fluids (water or liquid nitrogen, for instance). Three heating elements are inserted in holes symmetrically distributed around chamber center.

Different methods are employed for the thermal control of the chamber, depending on the temperature range. A thermostatic bath (TE 184, Tecnal) spans the range from 263 to 372 K by circulating water throughout the coil. Temperatures above 372 K are reached by tubular heating elements (NP 38899, HG Resistências). For temperatures below 263 K, liquid nitrogen vapor circulating in the coil is used for chamber cooling; heating elements are responsible for thermal stability in this case. The chamber and the coil are covered by aluminum foils during the experiments, in order to reduce heat changes with external environment. Tests indicate that the system can achieve temperature stability from 223 to 393 K.

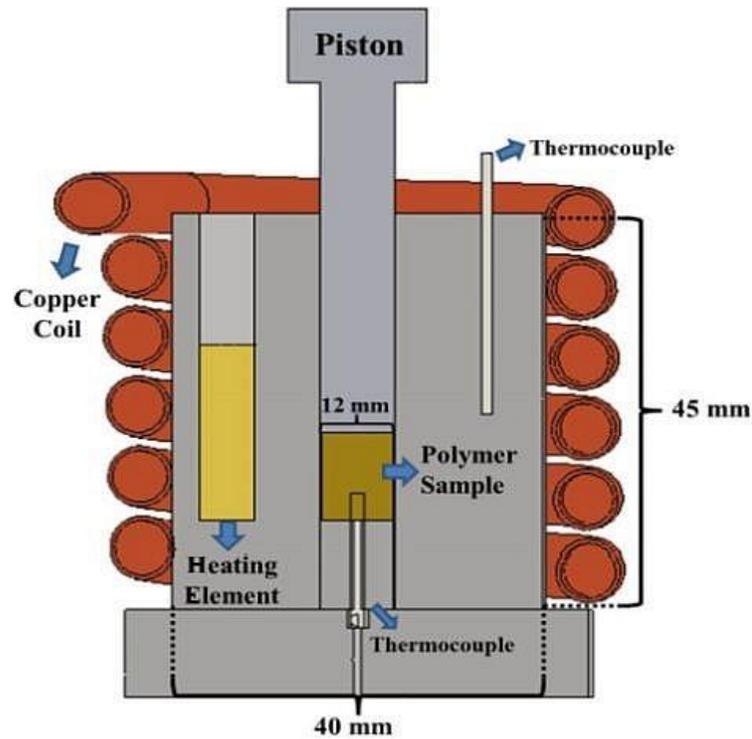

FIG 1: Schematic illustration of the cross-sectional view of the high-pressure chamber.

The independent setting of control parameters and the data acquisition from measuring instruments are implemented by a LabVIEW user interface. Thermal sensors consist of two thermocouples (type-K chromel-alumel, Omega Engineering): one stuck 5 mm inside the sample and another passing through a hole and reaching the middle of the chamber (see Fig. 1). The information obtained from the thermocouples are collected by a temperature controller (Model 335, Lake Shore Cryotronics), giving a temperature resolution of 0.01 K. This device also acts as PID controller of the heating elements, supplying and tuning the heat power during heating/cooling processes.

A load cell (3101C, ALFA Instrumentos) mounted under the set is used to measure the contact force, with 0.5 kgf of load resolution. The measured contact force is used to indirectly probe the sample pressure. The uniaxial load is applied by a 15,000-kgf hydraulic press (P15500, Bovenau). The manual control of the hydraulic ram allows load/unload increments within the resolution limit of the load cell. Nevertheless, additional uncertainties in load should be expected from the frictional contact between the chamber pieces and the sample. This issue is partially addressed by spreading a lubricant material over the piston and the chamber central hole. Considering the simple geometry of the pressure chamber and the direct sample-piston interaction, the pressure can be calculated by the straightforward relationship P = F / A, where F is the force measured by the load cell, and A is the cross-sectional area of the piston.

The sample strain is evaluated by a linear length gauge (METRO 2500, Heindenhain Co), which has 1 µm of resolution in displacement measurements. A metallic stripe, attached to the piston, leans on the length gauge. Through this stripe, the piston displacement is measured by the gauge as the loading/unloading process is performed. This quantity is directly related to the sample deformation (shrinkage or extension). The compressive strain variable used is this paper is $\varepsilon_c = (l-l_0) / l_0$, where $l_0$ is the initial length and $l$ is the current length. The capability of measuring the sample strain simultaneously with temperature and pressure represents a fundamental advantage of our apparatus over the barocaloric setups found in literature.[12–14]

**EXPERIMENTS WITH NR**

Two experiments were performed to verify the performance of the present experimental setup. The mechanocaloric material of choice in this study is NR, due its high caloric potential and resilience upon repetitive cycles of deformation. The NR cylindrical samples were synthesized from a pre-vulcanized latex resin (purchased from Siquiplas). Cylinders were shaped in a plaster mold during two days of natural cooling. The samples have diameter of 12 mm and density of 902(7) kg.m$^{-3}$; original length of $l_0$ = 19.5 mm. Figure 2 shows the typical time profile of ΔT, for the case of 1,000 kgf uniaxial load (compressive stress of 87(2) MPa) at room temperature (RT). Temperature (T) abruptly increases at $t_1$, as the sample is compressed. Then, the load is kept constant as the system exchanges heat with the surroundings; T decreases down to the initial value. At $t_2$, the load is removed rapidly, causing an abrupt decrease in T. Finally, the sample returns to RT as result of heat absorption.

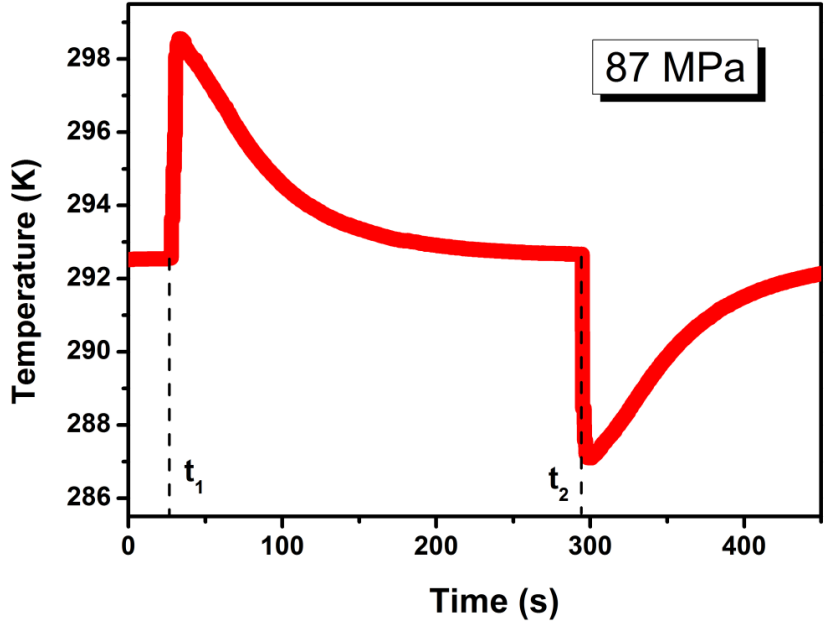

FIG 2: Typical temperature-time curve of NR under 87(2) MPa.

The experiment described above (Fig.2) was performed within the 283 − 333 K temperature range, at 43.4(9) and 87(2) MPa compressive stresses. Figure 3 shows the results for ΔT as function of the initial temperature. All curves exhibit a similar qualitative trend: as the initial temperature is raised, higher ΔT values are observed, increasing at an approximately constant rate. ΔT values in compression are slightly higher than in decompression for both pressures. The oscillation of ΔT data in compression at 43.4 MPa is related to the manual loading process, implying in different compression rates for each experiment, which affects the terminal temperature values. ΔT proportionally increases for measurements at 87 MPa in comparison with the 43.4 MPa curve. For all data sets, the measured ΔT's are comparable to those reported in $\sigma_e$-CE studies with NR, reaching ~ 6.5 K for 87 MPa (ref. 15 reports |ΔT| ~ 9 K for $\varepsilon = 6$).

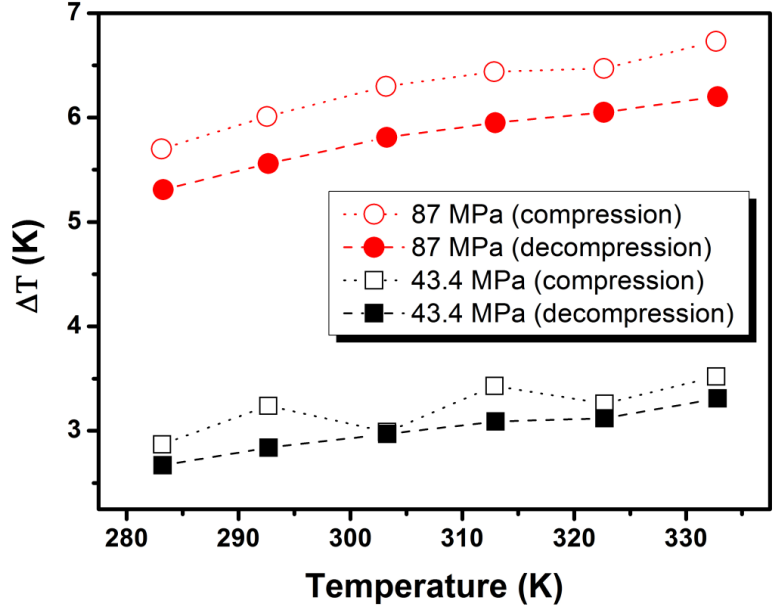

FIG 3: Measured temperature change as function of initial temperature.

The strain-temperature curves for NR were measured at different pressures, including the pressure of 26.0(6) MPa (Figure 4). Temperature was varied within the temperature range of 285 − 330 K. The data set reveals an inverse relationship between the strain and temperature, which is expected due the counter action of thermal expansion on the compression. The curves exhibit an approximately linear behavior in their central regions (317 − 294 K in cooling, 325 − 300 K in heating). We estimate systematic errors (an offset) of 0.002 and random errors of 0.001 for the strain values.

The strain–temperature curves allow to indirectly quantify the $\sigma_b$-CE by using the following Maxwell's relation[11,16,17]:

$$\Delta S_T(T, \Delta\sigma) = -\frac{1}{\rho_0} \int_{\sigma_1}^{\sigma_2} \left(\frac{\partial \varepsilon}{\partial T}\right)_\sigma d\sigma, \qquad (1)$$

where σ represents the compressive stress and $\rho_0$ the density of the sample at atmospheric pressure.

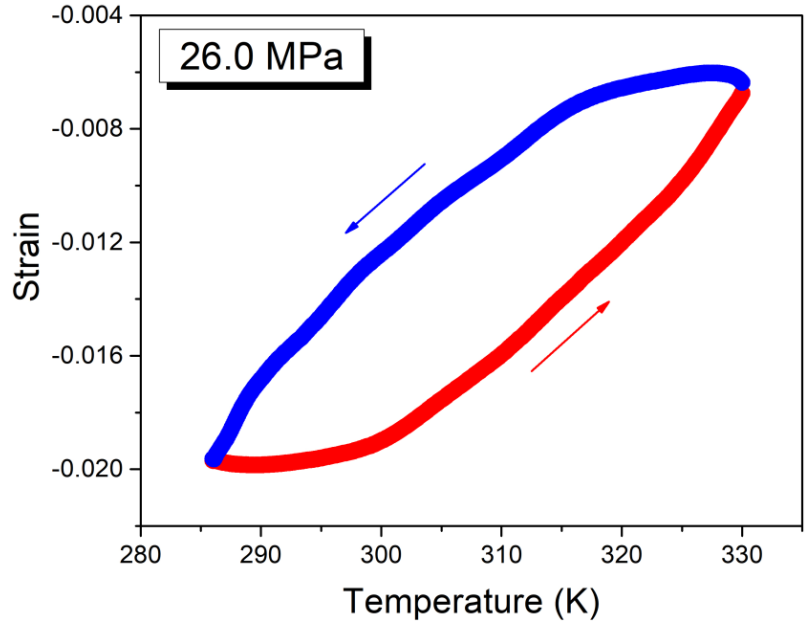

FIG 4: Strain-temperature curve of NR at constant pressure of 26.0(6) MPa.

**CONCLUSIONS**

We have presented a straightforward experimental setup to investigate the $\sigma_b$-CE in polymers. Thermal stability can be achieved within the $223 - 393$ K temperature range by means of heating elements and fluid circulation. The data collection system is composed by two thermocouples, a load cell and a linear length gauge, allowing measuring the relevant variables of the cooling cycle: temperature, pressure and compressive strain. The setup is fully functional for the proposed barocaloric experiments. Measured temperature changes are large and comparable to analogous studies with NR. Strain-temperature curves enable indirect quantifications of the isothermal entropy change through thermodynamic relations. The apparatus developed in this paper fulfills the gap of specific barocaloric studies concerning polymers, opening interesting perspectives for future mechanocaloric investigations for this class of materials.

**ACKNOWLEDGEMENTS**

We thank the technical support from João R. Costa, Orival da Silva, Fábio R. Zambello, Luis C. S. Vieira, and Adalberto F. M. Fontoura. We also thank the financial support from CNPq, CAPES, FAPESP (project number 2012/03480-0) and CNPEM.